\documentstyle[preprint,aps]{revtex}

\tightenlines

\begin{document}

\draft

\title{
A priori mixing of mesons and the ${\bf |\Delta I|=1/2}$ rule in 
${\bf K\rightarrow\pi\pi}$
}

\author{
A.~Garc\'{\i}a
}
\address{
Departamento de F\'{\i}sica.\\ 
Centro de Investigaci\'on y de Estudios Avanzados del IPN.\\
A.P. 14-740. M\'exico, D.F., 07000. MEXICO.
}
\author{ 
R.~Huerta
and
G.~S\'anchez-Col\'on
}
\address{
Departamento de F\'{\i}sica Aplicada.\\
Centro de Investigaci\'on y de Estudios Avanzados del IPN. Unidad Merida.\\ 
A.P. 73, Cordemex. M\'erida, Yucat\'an, 97310. MEXICO.
}

\date{\today}

\maketitle

\begin{abstract}
We consider the hypothesis of {\it a priori }mixings in the mass eigenstates of mesons to obtain the $|\Delta I|=1/2$ rule in $K\rightarrow\pi\pi$.
The Hamiltonian responsible for the transition is the strong interacting one.
The experimental data are described using the isospin symmetry relations between the strong coupling constants.
\end{abstract}


The explanation of the enhancement phenomenon observed in non-leptonic and weak radiative decays of hadrons (NLDH and WRDH respectively) has represented a challenge for already a long time~\cite{kagan}.
In previous work we have studied the possibility that this phenomenon be due,
not to some elaborate subtlety of strong interactions dressing weak vertices,
but to the existence of far away new physics leading to small admixtures of the
known hadrons with the new ones.
If these admixtures are flavor and parity violating they would contribute to observed NLDH and WRDH~\cite{apriori,detailed,universality}.
Because the weak interaction rates are so very much suppressed with respect to the strong and electromagnetic ones, it could well happen that, even if such
admixtures are very tiny, they might still give sizeable enough contributions
(through the strong and electromagnetic interaction Hamiltonians) as to exceed
$W$-mediated NLDH and WRDH and explain the enhancement phenomenon observed in
them.
We have refered to this scheme as ``a priori mixings" and in Refs.~\cite{apriori,detailed,universality} we have shown that indeed they can be used to describe very satisfactorily the experimental data of NLDH and WRDH.

Another notorious example of the enhancement phenomenon occurs in
$K\rightarrow \pi\pi$ decays.
If the a priori mixings approach is to be succesful it is imperative that it also describes these decays.
This is what we shall study in this paper.

The experimental data~\cite{pdg} on these decays can be presented in terms of the decay amplitudes $A_{+0}$, $A_{+-}$, and $A_{00}$, corresponding respectively to the modes $K^+\rightarrow\pi^+\pi^0$, $K_1\rightarrow\pi^+\pi^-$, and $K_1\rightarrow\pi^0\pi^0$, namely,

\[
|A_{+0}| = (1.831 \pm 0.006) \times 10^{-8} {\rm Gev},
\]

\begin{equation}
|A_{+-}| = (3.911 \pm 0.007) \times 10^{-7} {\rm Gev},
\label{seis}
\end{equation}

\[
|A_{00}| = (3.714 \pm 0.015) \times 10^{-7} {\rm Gev}.
\]

\noindent
One readily appreciates $|A_{+-}|\approx|A_{00}|\gg|A_{+0}|$.
In terms of isospin amplitudes of two weak interaction hamiltonians with $I=1/2$ and $I=3/2$, respectively, the experimental data mean that the $I=1/2$
piece is very importantly enhanced with respect to the $I=3/2$ one.
Actually, if the latter is neglected one should have $|A_{+-}|=|A_{00}|$ and $|A{+0}|=0$.
This has been refered to as the $|\Delta I|=1/2$ rule.

We shall now apply the a priori mixing scheme to the three decays.
The physical (mass eigenstates) mesons with parity and flavor violating admixtures are given by~\cite{detailed}

\[
K^+_{ph} =
K^+_{0p} -
\sigma \pi^+_{0p} -
\delta' \pi^+_{0s} +
\cdots
\]
 
\[
K^0_{ph} = 
K^0_{0p} +
\frac{1}{\sqrt 2} \sigma \pi^0_{0p} +
\sqrt{\frac{3}{2}} \sigma \eta_{0p} +
\sqrt{\frac{2}{3}} \delta \eta_{0s} +
\frac{1}{\sqrt 3} \delta \chi_{0s} +
\frac{1}{\sqrt 2} \delta' \pi^0_{0s} +
\frac{1}{\sqrt 6} \delta' \eta_{0s} -
\frac{1}{\sqrt 3} \delta' \chi_{0s} +
\cdots
\]

\[
\pi^+_{ph} = 
\pi^+_{0p} +
\sigma K^+_{0p} -
\delta K^+_{0s} +
\cdots
\]

\begin{equation}
\pi^0_{ph} =
\pi^0_{0p} -
\frac{1}{\sqrt{2}} \sigma (K^0_{0p} + \bar{K}^0_{0p}) +
\frac{1}{\sqrt{2}} \delta (K^0_{0s} - \bar{K}^0_{0s}) +
\cdots
\label{uno}
\end{equation}

\[
\pi^-_{ph} =
\pi^-_{0p} +
\sigma K^-_{0p} +
\delta K^-_{0s} +
\cdots
\]

\[
\bar{K}^0_{ph} =
\bar{K}^0_{0p} +
\frac{1}{\sqrt{2}} \sigma \pi^0_{0p} +
\sqrt{\frac{3}{2}} \sigma \eta_{0p} -
\sqrt{\frac{2}{3}} \delta \eta_{0s} -
\frac{1}{\sqrt{3}} \delta \chi_{0s} -
\frac{1}{\sqrt{2}} \delta' \pi^0_{0s} -
\frac{1}{\sqrt{6}} \delta' \eta_{0s} +
\frac{1}{\sqrt{3}} \delta' \chi_{0s} +
\cdots
\]

\[
K^-_{ph} =
K^-_{0p} -
\sigma \pi^-_{0p} +
\delta' \pi^-_{0s} +
\cdots
\]

\[
\eta_{ph} =
\eta_{0p} -
\sqrt{\frac{3}{2}} \sigma (K^0_{0p} + \bar{K}^0_{0p}) +
\frac{1}{\sqrt{6}} (\delta + 2\delta') (K^0_{0s} - \bar{K}^0_{0s}) +
\cdots
\]

\[
\chi_{ph} =
\chi_{0p} -
\frac{1}{\sqrt{3}} (\delta - \delta') (K^0_{0s} - \bar{K}^0_{0s}) +
\cdots.
\]

\noindent
In all these expressions the dots stand for other mixings that will not be used here, and the subindeces naught, $s$, and $p$ refer to strong-flavor, positive, and negative parity eigenstates.
Our phase conventions are those of Ref.\cite{gibson}.

Here, we shall not consider $CP$ violation and therefore the above states should not be necessarily $CP$-eigenstates, however notice that the physical mesons satisfy $CPK^+_{ph}=-K^-_{ph}$, etc.
We can form the $CP$-eigenstates $K_1$ and $K_2$ by

\begin{equation}
K_{1_{ph}} = \frac{1}{\sqrt{2}} (K^0_{ph} - \bar{K}^0_{ph})
\qquad\mbox{and}\qquad
K_{2_{ph}} = \frac{1}{\sqrt{2}} (K^0_{ph} + \bar{K}^0_{ph}),
\label{dos}
\end{equation}

\noindent
the $K_{1_{ph}}$ ($K_{2_{ph}}$) is an even (odd) state with respect to $CP$.

Substituing the expressions given in Eqs.~(\ref{uno}), we obtain (we droped
the naught subindex in the strong-flavor eigenstates to simplify the notation),

\[
K_{1_{ph}} =
K_{1_p} +
\frac{1}{\sqrt{3}} (2\delta + \delta') \eta_s +
\delta' \pi^0_s +
\sqrt{\frac{2}{3}} (\delta - \delta') \chi_s,
\]

\begin{equation}
K_{2_{ph}} =
K_{2_p} +
\sigma \pi^0_p +
\sqrt{3} \sigma \eta_p
\label{tres}
\end{equation}

\noindent
where the usual definitions
$K_{1_p} = (K^0_p - \bar{K}^0_p)/\sqrt{2}$
and
$K_{2_p} = (K^0_p + \bar{K}^0_p)/\sqrt{2}$
were used.

From Eqs.~(\ref{tres}) we obtain some important conclusions.
Since the Hamiltonian, $H_{st}$, responsible for the decay $K\rightarrow\pi\pi$ is by assumption isoscalar and also a flavour and parity conserving one, we notice that the physical state $K_{1_{ph}}$ can only decay into two pions and not into three pions.
In the latter case, the final state made out of three pions has total angular momentum equal to zero.
The parity is odd, since each of the pions has negative parity, and then the
Hamiltonian can not make the transition.
In the case of having two pions in the final state, the transition is possible and proportional to the constants $\delta$ and $\delta'$.
Similarly for the state $K_{2_{ph}}$: it has to go to three pions and the amplitude is proportional to $\sigma$.
The above qualitative behavior is observed experimentally neglecting $CP$-violation effects.
 
We now write explicitly the amplitudes for the decays $K\rightarrow\pi\pi$.
The three amplitudes we want are then,
$A_{+0} = \langle\pi^+_{ph}\pi^0_{ph}|H_{st}|K^+_{ph}\rangle$,
$A_{+-} = \langle\pi^+_{ph}\pi^-_{ph}|H_{st}|K_{1_{ph}}\rangle$,
and
$A_{00} = \langle\pi^0_{ph}\pi^0_{ph}|H_{st}|K_{1_{ph}}\rangle$.
After the substitution of the physical mass eigenstates given in Eqs.~(\ref{uno}) we obtain,

\[
A_{+0} =
-\delta'\langle\pi^+_p\pi^0_p|H_{st}|\pi^+_s\rangle +
\frac{1}{\sqrt{2}}\delta\langle\pi^+_p K^0_s|H_{st}|K^+_p\rangle
-\delta\langle K^+_s\pi^0_p|H_{st}|K^+_p\rangle
\]

\begin{eqnarray}
\label{cuatro}
A_{+-} &=&
\frac{1}{\sqrt{3}}(2\delta+\delta')\langle\pi^+_p\pi^-_p|H_{st}|\eta_s\rangle
+\delta'\langle\pi^+_p\pi^-_p|H_{st}|\pi^0_s\rangle
+\sqrt{\frac{2}{3}}(\delta-\delta')\langle\pi^+_p\pi^-_p|H_{st}|\chi_s\rangle
\nonumber\\
&&
-\frac{1}{\sqrt{2}}\delta\langle K^+_s\pi^-_p|H_{st}|K^0_{p}\rangle
-\frac{1}{\sqrt{2}}\delta\langle\pi^+_p K^-_s|H_{st}|\bar{K}^0_{p}\rangle
\end{eqnarray}

\begin{eqnarray*}
A_{00} &=&
\frac{1}{\sqrt{3}}(2\delta+\delta')\langle\pi^0_p\pi^0_p|H_{st}|\eta_s\rangle
+\delta'\langle\pi^0_p\pi^0_p|H_{st}|\pi^0_s\rangle
+\sqrt{\frac{2}{3}}(\delta-\delta')\langle\pi^0_p\pi^0_p|H_{st}|\chi_s\rangle
\\
&&
+\frac{1}{2}\delta\langle\pi^0_p K^0_s|H_{st}|K^0_p\rangle
+\frac{1}{2}\delta\langle\pi^0_p\bar{K}^0_s|H_{st}|\bar{K}^0_p\rangle
+\frac{1}{2}\delta\langle K^0_s\pi^0_p|H_{st}|K^0_p\rangle
+\frac{1}{2}\delta\langle\bar{K}^0_s\pi^0_p|H_{st}|\bar{K}^0_p\rangle
\end{eqnarray*}

\noindent
In the right hand side of these equations, the amplitudes are flavor and parity conserving.
The two pions in the physical final state of these amplitudes are either in the $I=0$ or in the $I=2$ isospin configuration, since the $I=1$ state is forbidden by the generalized Bose principle~\cite{neubert}.
Also, there is no contribution from the $I=2$ component since the interaction Hamiltonian is an isosinglet (so, amplitudes with a $\pi$ ($I=1$) in the initial state vanish).
Therefore, we can write such amplitudes in terms of a single strong coupling constant and take into account the final state interaction by introducing a multiplicative phase factor.
We will denote each amplitude in the form,
$\langle M^{1}_{i} M^{2}_{j}|H_{st}|M^{3}_{k}\rangle= G^{k,ij}_{M^3,M^1M^2}e^{i\alpha}$.

The amplitudes for the three decays considered become,

\[
A_{+0}= 
- \delta(\frac{1}{\sqrt{2}}G^{p,ps}_{K^+,\pi^+ K^0}
- G^{p,sp}_{K^+,K^+\pi^0})e^{i\alpha_{0}}
\]

\begin{equation}
A_{+-}=
[\frac{1}{\sqrt{3}}(2\delta + \delta')G^{s,pp}_{\eta,\pi^+\pi^-}
+ \sqrt{\frac{2}{3}}(\delta-\delta')G^{s,pp}_{\chi,\pi^+\pi^-}]e^{i\alpha_{1}}
\label{five}
\end{equation}

\[
A_{00}=
[\frac{1}{\sqrt{3}}(2\delta+\delta')G^{s,pp}_{\eta,\pi^0\pi^0}
+ \sqrt{\frac{2}{3}}(\delta-\delta')G^{s,pp}_{\chi,\pi^0\pi^0}]e^{i\alpha_{1}}
\]

\noindent
Above we have used the assumption that the strong coupling constants have the property,
$(G^{i,jk}_{M^3,M^1M^2})^{CPT}=
G^{i,jk}_{\overline{M}^3,\overline{M}^1 \overline{M}^2}$.
Also, we have used the properties,
$\langle j_1j_2m_1m_2|j_1j_2JM\rangle =
(-1)^{J-j_1-j_2}\langle j_2j_1m_2m_1|j_2j_1JM\rangle$,
of the $SU(2)$ Clebsch-Gordan coefficients to simplify the expressions for the amplitudes.
The phase introduced for the final state interaction depends only on the total isospin of the final particles, it is for this reason that $A_{+-}$ and $A_{00}$ have the same phase factor.

It is easy to see from Eqs.~(\ref{five}) that in the $SU(2)$ symmetry limit we obtain the so called $|\Delta I|=1/2$ rule predictions: $A_{+0} = 0$ and $A_{+-}=A_{00}$.
For instance, from the $SU(2)$ Clebsch-Gordan Tables we get
$G^{p,ps}_{K^+,\pi^+K^0} = \sqrt{2/3}G_{K,\pi K}$ and
$G^{p,sp}_{K^+,K^+\pi^0} = (-)(-1/\sqrt{3})G_{K,\pi K}$.

Since these $|\Delta I|=1/2$ rule equalities are not rigurously exact, it is important to remark that in the a priori mixing scheme their deviations are necessarily proportional to $SU(2)$ breaking contributions, in contrast to $W$-mediated decays where $SU(3)$ breaking is relevant (the mass of the $s$-quark).
That is, in the present scheme the following ratios are proportional to the order of the $SU(2)$ symmetry breaking, $\epsilon$,

\begin{equation}
\frac{|A_{+0}|}{|A_{+-}|} \approx
\frac{|A_{+0}|}{|A_{00}|} \approx
\frac{|A_{+-}-A_{00}|}{|A_{+-}|} \approx
\frac{|A_{+-}-A_{00}|}{|A_{00}|} \approx
\epsilon
\end{equation}

\noindent
From the experimental data of Eqs.~(\ref{seis}), these ratios are respectively

\begin{equation}
\label{eigth}
4.68 \% \approx
4.93 \% \approx
5.04 \% \approx
5.30 \% \approx
\epsilon
\end{equation}

\noindent
These numbers are indeed quite small and can be accepted as compatible with the order of $SU(2)$ breaking.
Actually, $\epsilon$ will become smaller if small contributions of the $W$-mediated decays are introduced, i.e. assuming that the $|\Delta I|=1/2$ piece of this hamiltonian is not enhanced and is of the size of the
$|\Delta I|=3/2$ piece.
In this case then one can conclude that the $SU(2)$ symmetry limit is even better than the estimate of Eq.~(\ref{eigth}).

Since the a priori mixing angles have been determined in Ref.~\cite{detailed} to be of order $10^{-7}$, we then see that the strong coupling constants (which will remain unmeasured for a long time) are of the order of one, as should be the case.
To close, let us finally mention that in the a priori mixing scheme the ``$|\Delta I|=1/2$" rule is really a ``$\Delta I = 0$" rule, because
the interaction hamiltonian is the strong flavour conserving one.

We would like to thank CONACyT (M\'exico) for partial support.

\end{document}